\documentclass[prl,twocolumn,aps,superscriptaddress,showpacs,floatfix]{revtex4}

\usepackage{graphicx}
\usepackage{dcolumn}
\usepackage{bm}

\begin{document}
\title{Geometric Stochastic Resonance}

\author{Pulak Kumar Ghosh}
\affiliation{Advanced Science Institute, The Institute of Physical
and Chemical Research (RIKEN), Wako-shi, Saitama, 351-0198, Japan}
\author{Fabio Marchesoni}
\affiliation{Advanced Science Institute, The Institute of Physical
and Chemical Research (RIKEN), Wako-shi, Saitama, 351-0198, Japan}
\affiliation{Dipartimento di Fisica, Universit\`a di Camerino,
I-62032 Camerino, Italy}
\author{Sergey E. Savel'ev}
\affiliation{Advanced Science Institute, The Institute of Physical
and Chemical Research (RIKEN), Wako-shi, Saitama, 351-0198, Japan}
\affiliation{Department of Physics, Loughborough University,
Loughborough LE11 3TU, United Kingdom}

\author{Franco Nori}
\affiliation{Advanced Science Institute, The Institute of Physical
and Chemical Research (RIKEN), Wako-shi, Saitama, 351-0198, Japan}
\affiliation{Department of Physics, University of Michigan, Ann
Arbor, MI 48109-1040, USA}
\date{\today}
\begin{abstract}
A Brownian particle moving across a porous membrane subject to an
oscillating force exhibits stochastic resonance with properties
which strongly depend on the geometry of the confining cavities on
the two sides of the membrane. Such a manifestation of stochastic
resonance requires neither energetic nor entropic barriers, and can
thus be regarded as a purely geometric effect. The magnitude of this
effect is sensitive to the geometry of both the cavities and the
pores, thus leading to distinctive optimal synchronization
conditions.
\end{abstract}

 \maketitle

Stochastic resonance (SR) is by now a textbook example of how noise
can best enhance the response of a bistable system to an external
drive \cite{RMP}. Historically the research on SR focused mostly on
systems with purely energetic potentials, either continuous or
discrete. However, as pointed out in Ref.~\cite{Reguera}, in soft
condensed matter and in a variety of biological systems
\cite{natural}, particles are often confined to constrained
geometries, such as interstices, pores, or channels, whose size and
shape can affect the SR mechanism \cite{Burada1}. Indeed, smooth
confining geometries can be modeled as entropic (i.e., noise or
temperature dependent) potentials \cite{Zwanzig}, capable of
influencing the response of the system to an external driving force
(see, for a review, Ref. \cite{chemphyschem}).

\begin{figure}[tp]
\centering
\includegraphics[width=8.5cm]{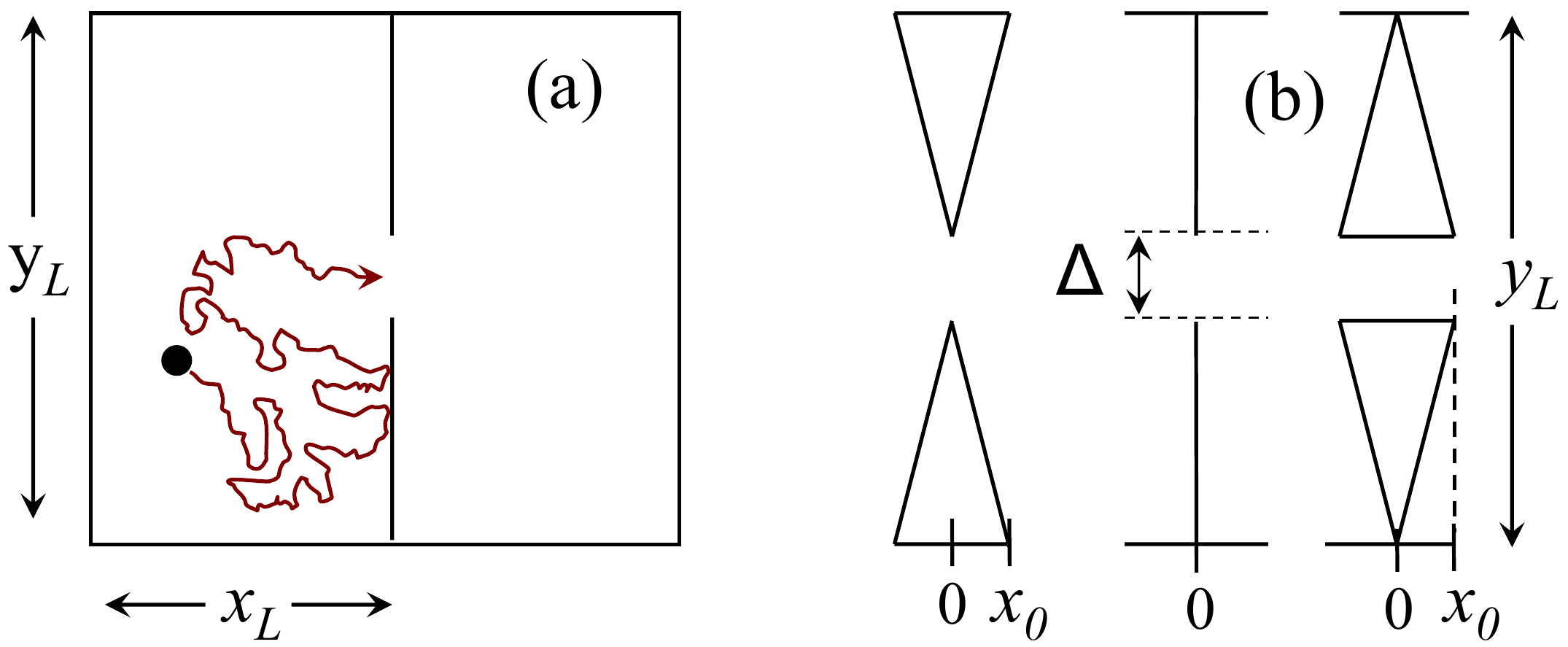}
\caption{(Color online) (a) Brownian particle confined to a 2D box
divided in two compartments by a partition with an opening at the
center. (b) Three different pore geometries used in our simulations:
funnel, hole, and spout (left to right). $\Delta$ and $x_0$ denote,
respectively, the cross-section and the thickness of the pore.
\label{F1}}
\end{figure}

Let us consider a Brownian particle freely diffusing in a two (2D)
or three dimensional suspension fluid contained in two symmetric
cavities connected by a narrow pore; it can switch cavity only by
overcoming the entropic barrier determined by the geometric
constriction associated with the pore. This is true even for ideal
reflecting boundaries, that is, in the absence of an intrinsic
energetic barrier. Burada {\it et al.} \cite{Burada1} have shown
that entropic barriers significantly contribute to the magnitude of
the SR effect that occurs when a periodic force drives the particle
across the pore; hence the term of ``entropic SR" coined in Ref.
\cite{Burada1}. The evidence reported there, however, hints at an
interplay of entropic and energetic barriers, rather than to a mere
entropic effect. Indeed, SR was demonstrated there only under the
explicit condition that the applied field of force had a dc
component at an angle with the pore axis, say orthogonal to it, so
that the particle tended to sojourn preferably against one side of
the cavities. In the absence of such an additional symmetry-breaking
force, no SR was
observed. 

This remark raises the issue whether a bistable effective potential
is a necessary condition for SR to occur altogether \cite{Borromeo}.
In this paper we show that a Brownian particle confined to two
distinct cavities divided by a porous medium, say a membrane, does
undergo SR when driven by an ac force perpendicular to the membrane,
even in the absence of external gradients and/or interactions with
the walls (besides bouncing from the walls). At variance with
ordinary SR, optimal synchronization between drive and particle
oscillations for an appropriate noise level, {\it only} occurs at
driving frequencies (amplitudes) lower (higher) than a certain onset
threshold. Moreover, such a manifestation of SR in higher dimensions
requires adopting extremely sharp geometrical constrictions to
separate the two cavities, something akin to the pores obtained by
puncturing a thin membrane. The magnitude and conditions of the
effect reported here are sensitive to both the geometry of the
cavities and the cross-section of the pores, thus allowing a direct
control of the synchronization mechanism.

\begin{figure}[tp]
\centering
\includegraphics[width=8cm]{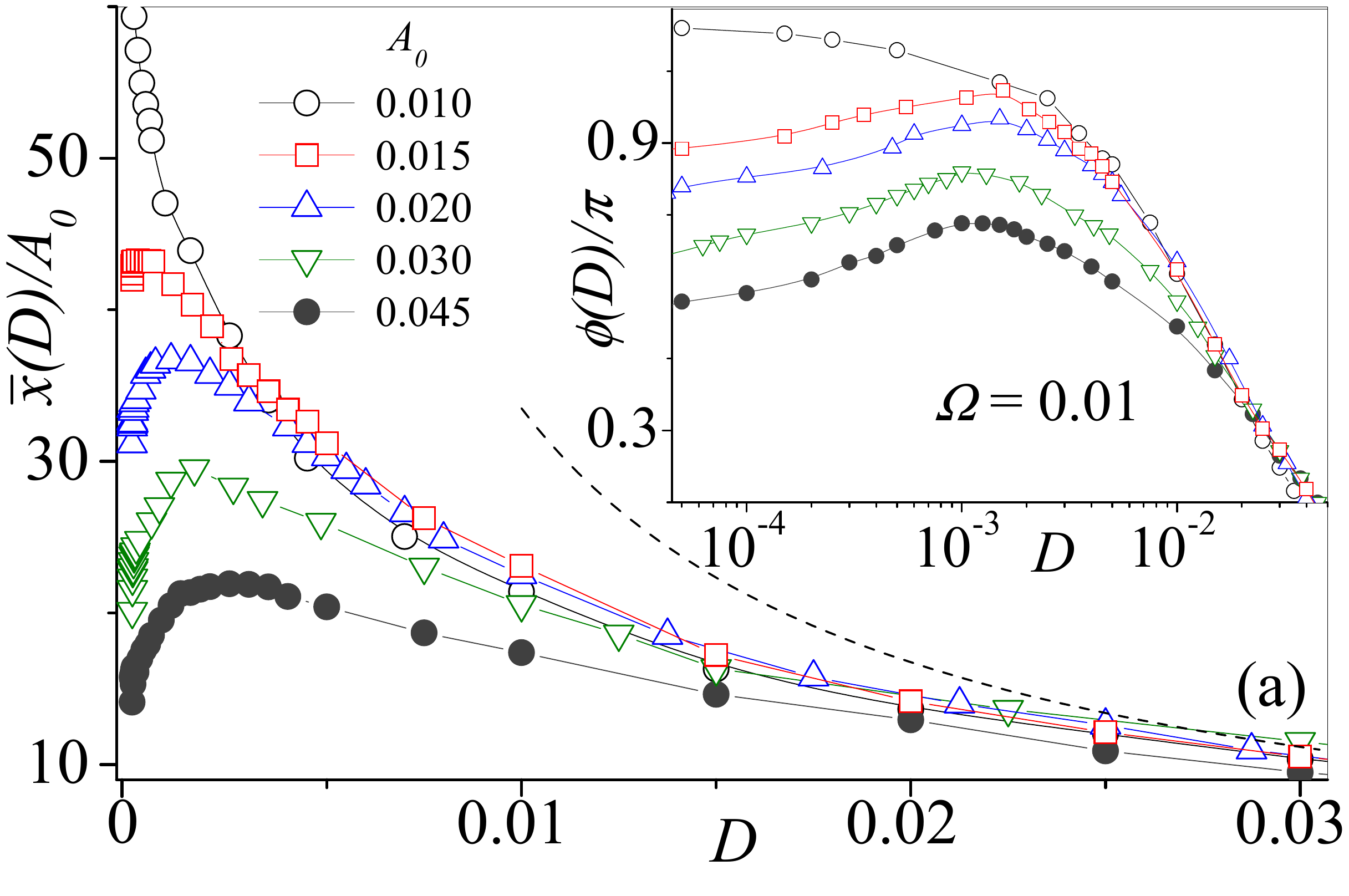}
\includegraphics[width=7.8cm]{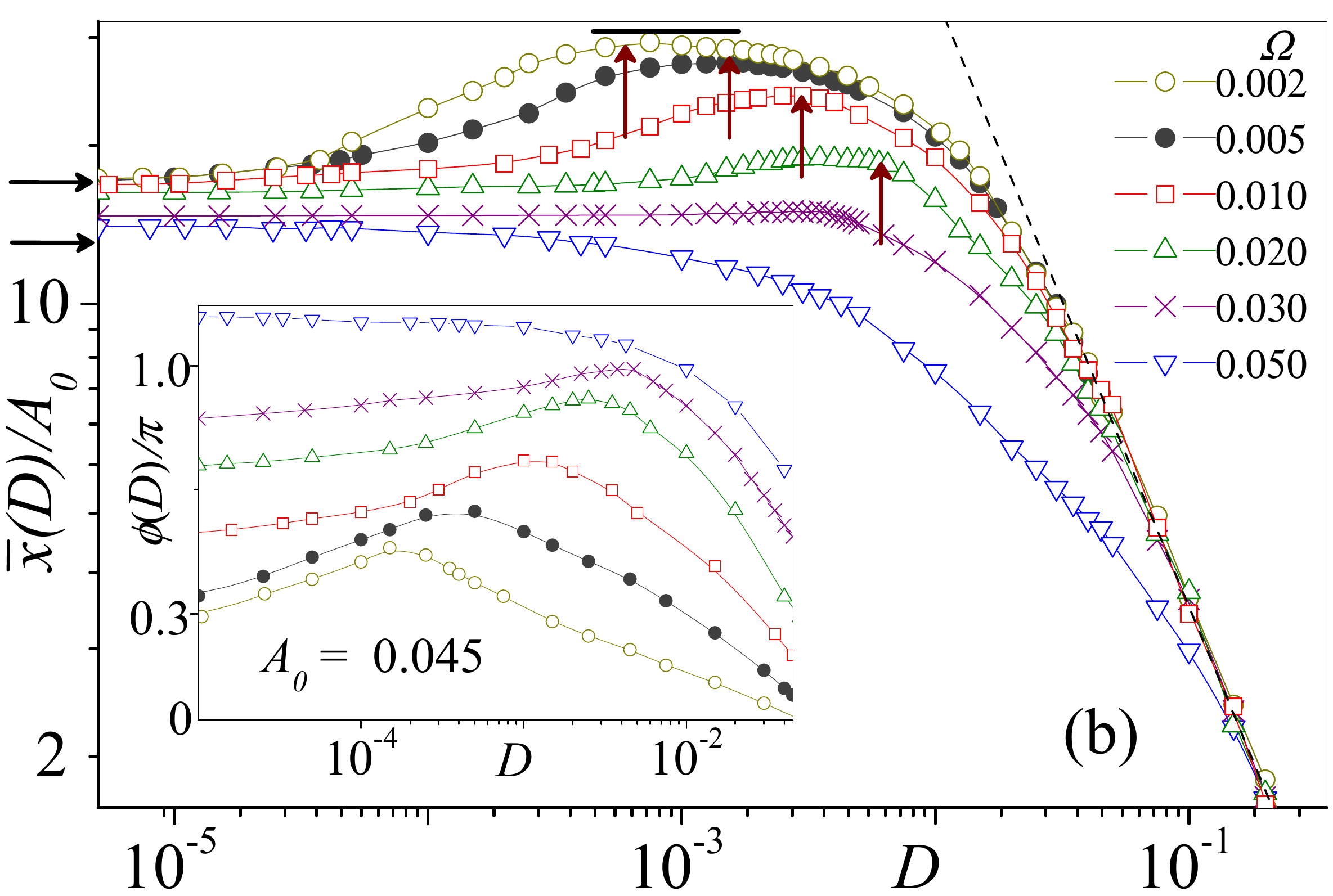}
\caption{(Color online) Geometric stochastic resonance. (a)
$\overline x(D)$ versus $D$ for different values of the ac drive
parameters $A_0$ at $\Omega=0.01$ in (a) and $\Omega$ at $A_0=0.045$
in (b). Other parameters are: $x_L=y_L=1$ and $\Delta =0.1$. The
dashed curve in (a) represents our predicted asymptotic decay
$\overline x/A_0$ for $D \to \infty$.  In (b) we display our
predictions for: the range of variability of $\overline x(0)$
(horizontal arrows), the SR peak position $D_{\rm max}$ (vertical
arrows), the SR peak height $\overline x(D_{\rm max})$ (top line),
and the decay law $\overline x(D\to \infty)$ (dashed line); see
text.\label{F2}}
\end{figure}

\begin{figure}[tp]
\centering
\includegraphics[width=8.2cm]{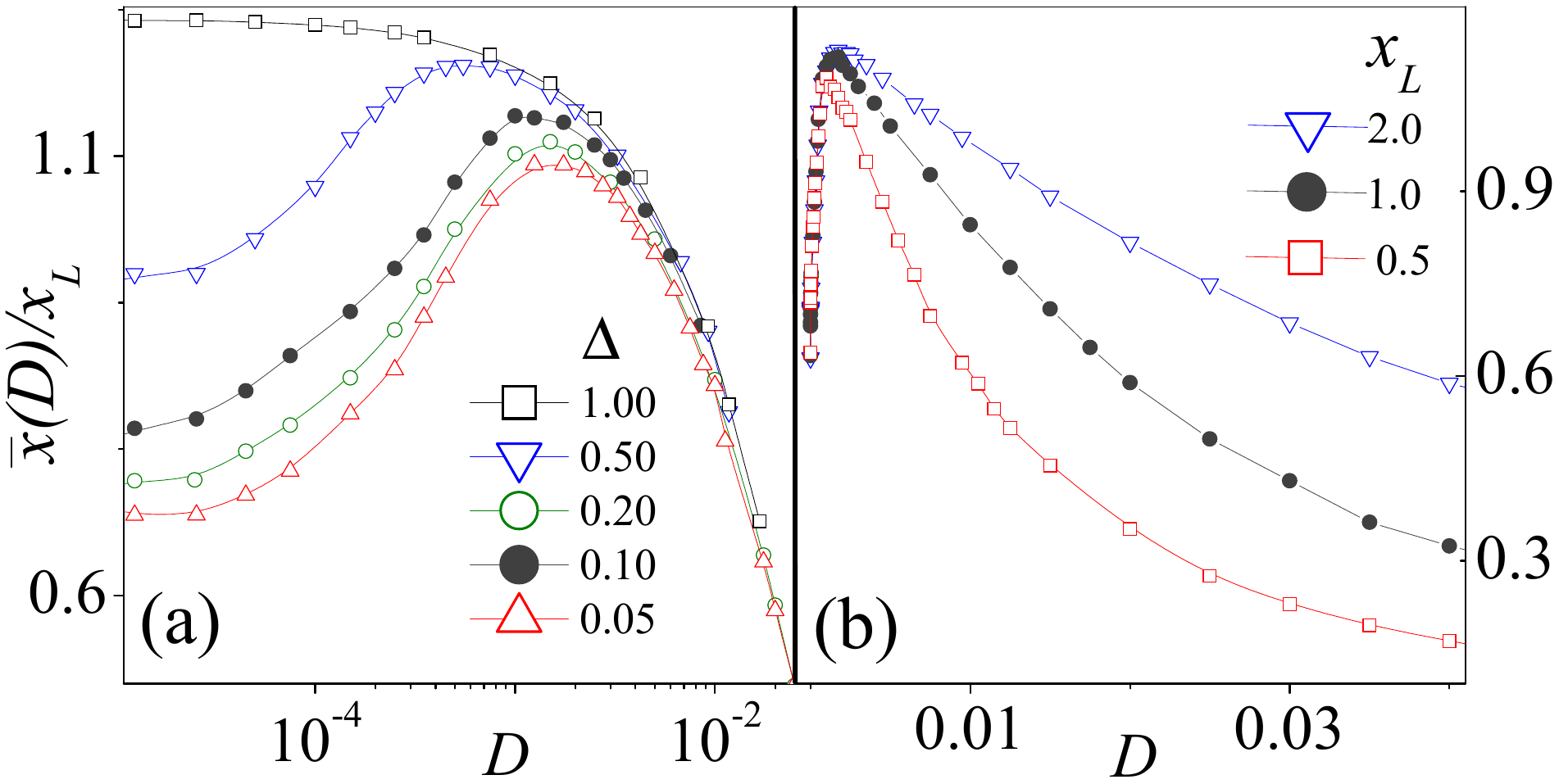}
\caption{(Color online) Geometry dependence of this SR. $\overline
x(D)$ versus $D$ for different values of $\Delta$ (a) and $x_L$ (b).
Other parameters: $y_L=1$, $A_0=0.045$, $\Omega =0.005$ and (a)
$x_L=1$, (b) $\Delta=0.1$. In both panels $\overline x(0)$ is very
close to our prediction of Eq.~(\ref{xzero}) with
$\kappa(\Omega)=4/\pi$. \label{F3}}
\end{figure}

\begin{figure}[tp]
\centering
\includegraphics[width=8.0cm]{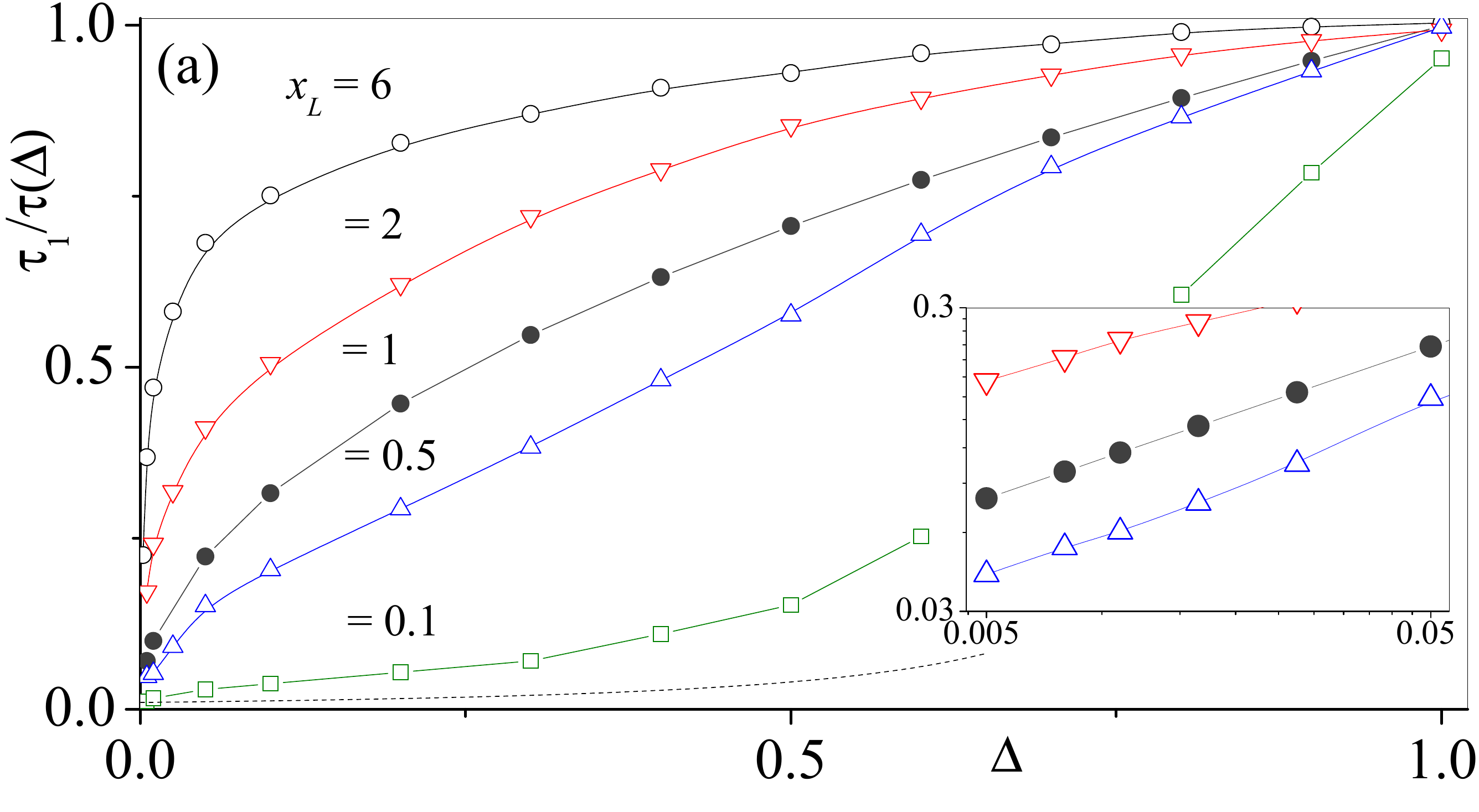}
\includegraphics[width=8.0cm]{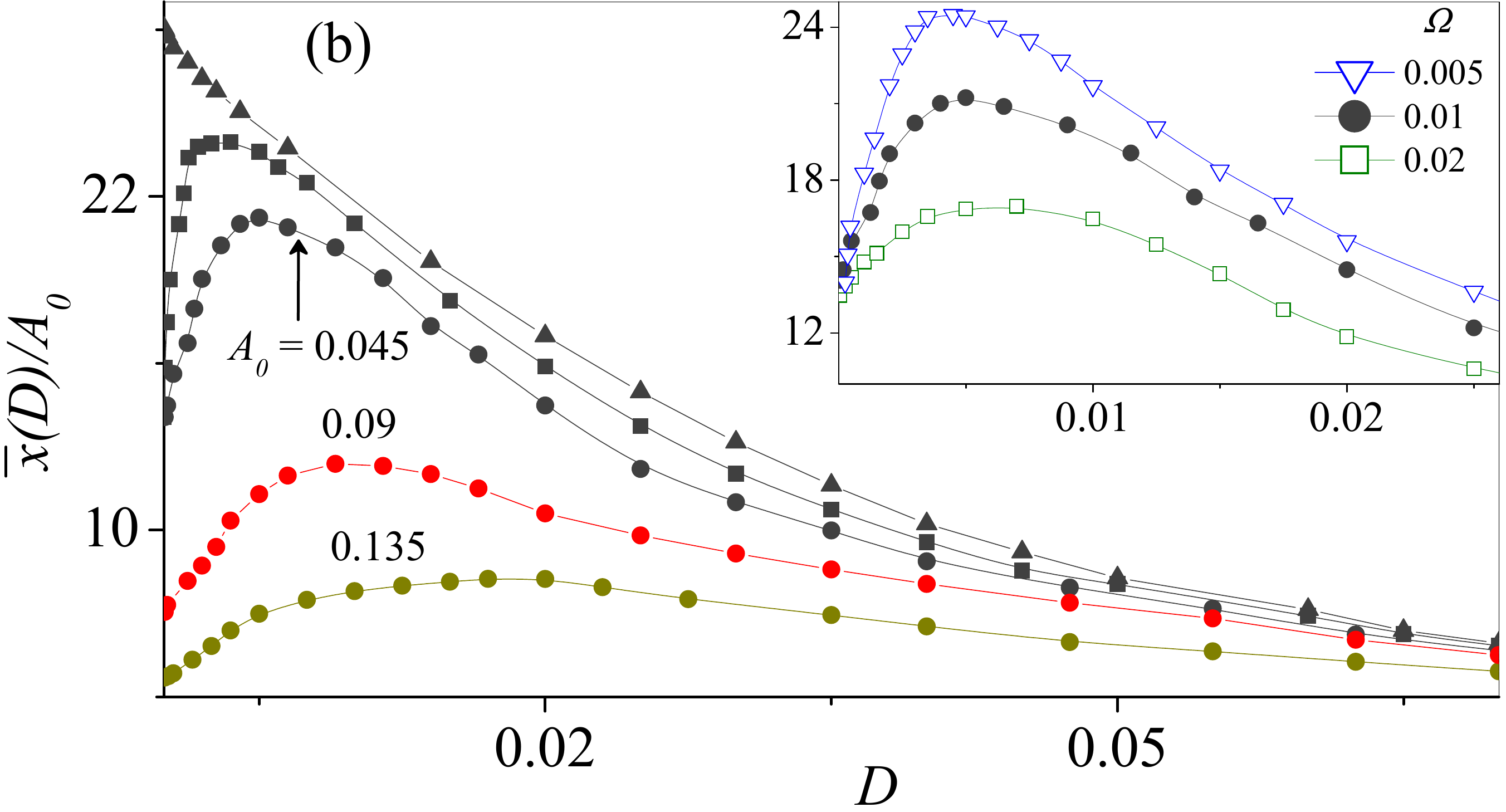}
\caption{(Color online) Dependence of SR on the pore geometry. (a)
$\tau(\Delta)$ in units of $\tau_1$ for different cavity lengths.
The drive has been switched off and the initial position of the
particle uniformly distributed within one cavity; averages were
taken over $2\times 10^5$ trajectories. The dashed curve represents
the vertical diffusion time, $(y_L-\Delta)^2/12D$ (see text), for
the shortest cavity. Inset: scaling law $\tau_1/\tau(\Delta) \propto
(y_L/\Delta)^{1/2}$ for $\Delta \to 0$. (b) $\overline x(D)$ versus
$D$ for different pore geometries: spout (circles), hole (squares),
and funnel (triangles). $A(t)$ is a square waveform with $\Omega
=0.01$ and amplitude as in the legend. Inset: $\overline x(D)$
versus $D$ for a spout-like pore at $A_0=0.045$ and different
$\Omega$. Other simulation parameters are: $x_L=y_L=1$,
$\Delta=0.1$, and $x_0=0.1$ (where it applies).\label{F4}}
\end{figure}

The overdamped dynamics of a Brownian particle in 2D is modeled by
the Langevin equation
\begin{equation}\label{le}
d{\vec r}/{dt}=-A(t)\;{\vec e}_x + \sqrt{D}~{\vec \xi}(t),
\end{equation}
where ${\vec e}_x, {\vec e}_y$ are the unit vectors along the $x, y$
axes and ${\vec \xi}(t)=(\xi_x(t),\xi_y(t))$ are zero-mean, white
Gaussian noises with autocorrelation functions $\langle
\xi_i(t)\xi_j(t')\rangle = 2\delta_{ij}\delta(t-t')$ with $i,j=x,y$.
Equation (\ref{le}) has been numerically integrated for the
two-cavity container sketched in Fig.~\ref{F1}, with reflecting
walls \cite{gardiner} and a single opening, with different
geometries, placed at the center of the partition wall. In the
presence of an ac drive, $A(t)=A_0\cos (\Omega t)$, the Brownian
trajectories embed a persistent harmonic component, $\overline x(D)
\cos[\Omega t -\phi(D)]$, whose amplitude, $\overline x$, and phase,
$\phi$, are plotted versus $D$ in Figs.~\ref{F2}-\ref{F4}.

For simplicity, we start modeling the pore as a structureless hole
pierced in a zero-thickness wall. The occurrence of a SR phenomenon
is clear, albeit with some distinctive features, as seen in
Figs.~\ref{F2} and \ref{F3}: (1) $\overline x(D)$ peaks for an
appropriate noise intensity, $D_{\rm max}$, with $D_{\rm max}$ a
manifestly increasing function of $\Omega$; (2) SR is restricted to
$A_0>A_c$, Fig.~\ref{F2}(a), and $\Omega < \Omega_c$,
Fig.~\ref{F2}(b). This is an important difference with respect to
ordinary SR, where there exist no such onset thresholds in the drive
parameters space; (3) The curves $\overline x(D)$ decay like
$D^{-1}$, Fig.~\ref{F3}, that is faster than in any 1D bistable
potential and more in line with SR in a discrete two-state model
\cite{heinsalu}.

All properties listed above can be explained by simple geometrical
considerations. We start noticing that when the ac force $A(t)$
presses the particle against the walls of the container opposite to
the dividing wall, then the average particle displacement $\langle x
(t) \rangle$ approaches a square waveform with amplitude $x_L$.
Correspondingly, the particle gets pushed against the container
partition twice per cycle; if it goes through the opening, $\langle
x (t) \rangle$ traces a symmetric, two-sided square wave; if it
doesn't, its average displacement is restricted to an asymmetric,
one-sided square wave, on either the positive or the negative side
of the dividing wall. On taking the Fourier series of $\langle x (t)
\rangle$ with period $2\pi/\Omega$, the amplitude of its fundamental
harmonic component turns out to be, respectively, $4x_L/\pi$ for the
two-sided waveform, and $2x_L/\pi$ for the one-sided waveform. From
this remark there follow immediately an upper bound to $\overline
x$,
\begin{equation}\label{xmax}
\overline x \leq ({4}/{\pi})~x_L,
\end{equation}
and the onset condition for SR,
\begin{equation}\label{onset}
{A_0}/{\Omega} \geq ({4}/{\pi})~x_L,
\end{equation}
$A_0/\Omega$ being the driven oscillation amplitude of an
unconstrained Brownian particle. The SR peak, $\overline x(D_{\rm
max})$, approaches the upper bound in Eq.~(\ref{xmax}) at
vanishingly low $\Omega$, see Fig.~\ref{F2}(b); whereas from
Eq.~(\ref{onset}), for our simulation parameters one obtains
$A_c=0.013$ in Fig.~\ref{F2}(a) and $\Omega_c=0.035$ in
Fig.~\ref{F2}(b). The consistence of these analytical results with
the simulations is quite satisfactory.

At variance with SR in a bistable potential, the particle
oscillations are not drastically suppressed in the zero-noise limit.
An estimate for $\overline x(0)\equiv \lim_{D\to 0} \overline x(D)$
can be obtained by noticing that the probability for the particle to
cross the pore and, therefore, to execute a full oscillation is
$\Delta/y_L$, whereas the probability to get trapped on either side
of the partition is $(1-\Delta/y_L)$. Accordingly,
\begin{equation}\label{xzero}
\overline x(0)=(x_L/2)(1+\Delta/y_L)\kappa(\Omega),
\end{equation}
with $\kappa=4/\pi$ for $\Omega \to 0$, see Eq.~(\ref{xmax}), and
$\kappa = 1$ for $\Omega \to \Omega_c$. This defines a relatively
narrow variability range for $\overline x(0)$ as a function of
$\Omega$. Well above the onset threshold (\ref{onset}), namely for
$\Omega \ll \Omega_c$, for $D\to 0$ the curves in Fig.~\ref{F2}(b)
clearly tend to the upper bound of $\overline x(0)$ in
Eq.~(\ref{xzero}); correspondingly, the upper bound of noise
amplification via SR is $\overline x(D)/\overline x(0)\leq 0.5$.

To check these predictions we modified the container geometry. In
Fig.~\ref{F3} the response amplitude $\overline x(D)$, in units of
$x_L$, is plotted versus $D$ for different $\Delta$'s, panel (a),
and $x_L$'s, panel (b). In agreement with Eq.~(\ref{xzero}),
$\overline x(0)/x_L$ in panel (a) grows linearly with $\Delta$,
while in panel (b) is seemingly insensitive to $x_L$. Note that
$\overline x(0)$ approaches the upper bound (\ref{xmax}) only for
$\Delta \to 1$, thus implying that geometric SR happens for any
finite pore width.

The dependence of the decaying tails of $\overline x(D)$ on $\Delta$
and $x_L$ is also consistent with our geometric interpretation.
Indeed, the $D \to \infty$ behavior of $\overline x(D)$ can be
analyzed in terms of the {\it average} time, $\tau_1(D)$, for an
unbiased Brownian particle to diffuse across one compartment. Such
time constant is easily obtainable by analytical means
\cite{gardiner}, $\tau_1(D)=x_L^2/3D$. In the presence of strong
noise, the geometric constriction exerted by the pore grows
ineffective; diffusion along the $x$-axis is then described by an ac
forced damped Brownian motion with effective damping constant
$\tau_1^{-1}$, so that the corresponding $\Omega$ component of
$\langle x(t) \rangle$ gets suppressed both in amplitude,
$A_0\tau_1(D)/\sqrt{1+ [\Omega\tau_1(D)]^2}$, and phase,
$\phi(D)=\arctan[\Omega\tau_1(D)]$. For $\Omega \ll \Omega_c$, this
leads to the $\Omega$-independent estimate, $\overline x(D)\sim
A_0\tau_1(D)$, plotted in both panels of Fig.~\ref{F2} (dashed
curves). Note that $\tau_1$ is independent of $\Delta$ and quadratic
in $x_L$, so that for $D\to \infty$ the tails of the curves of
Fig.~\ref{F3}(a) collapse onto one curve and those of
Fig.~\ref{F3}(b) scale proportional to $x_L$.

The SR peaks of $\overline x(D)$ occur at a certain value of the
noise strength, $D_{\rm max}$, see Fig.~\ref{F2}, which weakly
depends on $A_0$ and grows with $\Omega$. This property
distinguishes geometric SR from ordinary SR, as here the $x(t)$
transitions between stable states (cavities) are not regulated by an
activation rate over an energetic barrier of the Arrhenius type. In
the absence of a drive, $A_0=0$, the relaxation is characterized by
some exit time, $\tau(\Delta)$, which is necessarily inverse
proportional to $D$. Optimal synchronization between pore crossings
and external drive requires that the particle switches compartment
twice during one $A(t)$ cycle \cite{RMP}, namely
\begin{equation}\label{SRpeak}
\tau(\Delta)=\pi/\Omega.
\end{equation}
As a consequence, $D_{\rm max}$ is proportional to $\Omega$.

For a quantitative analysis, we propose to define $\tau(\Delta)$ as
the mean time a Brownian particle, uniformly distributed in one
compartment, first crosses the pore. This quantity has been
numerically computed and plotted in Fig.~\ref{F4}(a) for $A_0=0$ and
different container geometries. On inserting the corresponding
numerical data for $\tau(\Delta)$ into Eq.~(\ref{SRpeak}), we
predicted the values of $D_{\rm max}$ marked in Fig.~\ref{F2}(b) by
vertical arrows.

The dependence of $\tau(\Delta)$ on the parameters $\Delta$ and
$x_L$ is also instructive. For $\Delta=y_L$ we recovered the 1D
limit, $\tau_1$, as it should.  At very small pore cross-sections,
$\tau(\Delta)$ turned out to scale like $(y_L/\Delta)^{1/2}$, no
matter what the compartment aspect ratio $x_L/y_L$. However, the
ratio $x_L/y_L$ controls the interplay between diffusion along the
$x$ and $y$ axis. For long compartments, $x_L/y_L \gg 1$, the exit
time is of the order of $\tau_1$ until quite low $\Delta$, whereas
for narrow compartments, $x_L/y_L \ll 1$, the exit time is dominated
by the vertical diffusion of the particle towards the mid-section of
the cavity, where the pore is located, i.e., $\tau(\Delta)\sim
(y_L-\Delta)^2/12D$ for $\Delta$ not too close to 0 and 1 [also
shown in Fig.~\ref{F4}(a)]. To this regard, we remark that such a
dependence of the exit time on $x_L/y_L$ is not inconsistent {\it
per se} with a reduced 1D description of the Brownian diffusion
along the $x$-axis. However, {\it none} of the assumptions
introduced in Ref.~\cite{Burada1} to accommodate for SR in the
framework of the Fick-Jacobs kinetics \cite{Jacobs}, apply to the
present case. In particular, {\it no} entropic barrier can be
naturally attributed to the pore. The mechanism investigated here is
rather an irreducible 2D geometric effect, for which the term {\it
geometric SR} is more appropriate.

The phase delay, $\phi(D)$, of the fundamental harmonic component of
$\langle x(t) \rangle$ with respect to $A(t)$ is plotted in the
insets of Fig.~\ref{F2}. By comparing the corresponding curves for
$\overline x(D)$ and $\phi(D)$, we notice that: (1) $\phi(D)$ also
exhibit SR-like peaks, which shift to higher $D$ on increasing
$\Omega$ [Fig.~\ref{F2}(b)], while being almost insensitive to $A_0$
[Fig.~\ref{F2}(a)]. Note that in a two-state model $\phi(D)$ would
be a monotonically decreasing function of $D$ \cite{RMP}. (2) The
phase peaks get suppressed at low $\Omega$, where the SR signature
of $\overline x(D)$ is the most prominent; (3) In the zero noise
limit, contrary to $\overline x(0)$, $\phi(0)$ strongly depends on
$\Omega$ and $A_0$; for $D\to \infty$ the expected asymptotic
behavior, $\phi(D)\sim \Omega \tau_1(D)$, was recovered (not shown).
The nonmonotonic behavior of $\phi(D)$ signals the appearance of
two-sided oscillating trajectories in the averaging ensemble of
$\langle x(t) \rangle$, with the particle crossing more and more
frequently the pore in unison with $A(t)$. According to our
interpretation of the phenomenon under study, phase peaks are
natural SR precursors. However, if the SR onset condition
(\ref{onset}) is not met, $\phi(D)$ tends to $\pi/2$, as expected,
as the particle is restricted to oscillate inside one cavity most of
the time \cite{RMP}.

In real experiments at small-length scales, the geometry of the
partition wall and its opening(s) are often not fully controllable.
In Fig.~\ref{F4}(b) we plot $\overline x(D)$ versus $D$ for the
three different pore shapes sketched in Fig.~\ref{F1}(b). In order
to enhance geometry effects, we simulated a square waveform $A(t)$
also with amplitude $A_0$ and angular frequency $\Omega$. Moreover,
the pore width, $x_0$, was taken not too small, lest the different
pore geometries become indistinguishable. Funnel-like pores tend to
channel the Brownian trajectories through the opening, no matter
what their cross-section; thus, no SR evidence was detected, as
anticipated in Ref.~\cite{Burada1}. On the contrary, spout-like
pores exhibit enhanced SR peaks and a distinct dependence on the
cavity geometry. Indeed, under the pressure exerted by the drive,
the exit time through a spout of width $x_0$ is of Arrhenius type,
namely, a function of the pore cross-section, proportional to
$e^{A_0x_0/D}$.
Owing to the SR condition (\ref{SRpeak}), $D_{\rm max}$ is now
expected to grow almost linearly with $A_0$ and logarithmically
decreases with $\Omega$, in qualitative agreement with our
simulations. Note that, as soon as the pore shape comes into play,
i.e., for $x_0A_0\gg D$, the $A_0$ dependence of the exit time
cannot be ignored any more and a quantitative analysis of these
results requires going beyond the approximations of the linear
response theory \cite{RMP}. In the opposite limit, the pore can be
well modeled by a simple hole, as we initially did.

We expect that geometric SR can be best demonstrated in vortex
superconducting devices \cite{vortex}. This class of artificial
devices is presently attracting growing interest because of
potential applications to flux qubits, SQUIDs and superconducting rf
filters. Ion-beam techniques allow to fabricate a superconducting
sample with two vortex boxes connected by a thin pore of almost any
geometry. Vortices are trapped inside the boxes with binding energy
of the order of $\Phi_0^2L_t/\lambda^2$, where $\Phi_0$ is the
magnetic flux quantum, $\lambda$ is the London penetration depth,
and $L_t$ is the depth of the two vortex traps. The vortex density
$n=H/\Phi_0$, is controlled by the intensity $H$ of the applied
magnetic field. In the dilute limit, $H\lesssim \Phi_0/\lambda^2$,
the vortex-vortex interactions become negligible, so that the
transport properties of a single trapped vortex are not overshadowed
by many-body effects. ac drives and noise sources can be easily
implemented as Lorentz forces generated by independent electric
currents injected into the sample parallel and perpendicular to the
pore axis. Detection of SR under such experimental conditions is
regulated by the applied current sources only; in particular, the
noise parameter $D$ can be varied independently of the constant
operating sample temperature.

We thank the RIKEN Super Combined Cluster System for providing us
with extensive computing resources. We acknowledge partial support
from the NSA, LPS, ARO, NSF Grant No. EIA-0130383, the EPSRC Grants
No. EP/ D072581/1 and No. EP/F005482/1, and the AQDJJ.

\end{document}